\documentclass{ws-p8-50x6-00}

\def\p{\partial}
\def\r{\rho}
\def\half{{\scriptstyle\frac{1}{2}}}

\def\PRA{{\it Phys. Rev.\/} A}

\begin{document}

\title{On hybrid dynamics of the Copenhagen dichotomic world}

\author{Lajos Di\'osi}

\address{Institute for Advanced Study, Wallotstrasse 19, D-14193 Berlin,
Germany}

\address{Research Institute for Particle and Nuclear Physics, H-1525
Budapest 114, POB 49, Hungary\\
E-mail: diosi@rmki.kfki.hu}

\maketitle

\abstracts{In the Copenhagen viewpoint, part of the world is quantized and the 
complementary part remains classical. From a formal dynamic aspect, 
standard theory is incomplete since it does never account for the so-called
'back-reaction' of quantized systems on classical systems except for the highly
idealized system--detector interaction. To resolve this formal issue, a 
certain 'hybrid dynamics' can be constructed to account for the generic 
interaction between classical and quantized parts. Hybrid dynamics incorporates
 standard quantum theory, including collapse of the wave function during 
system--detector interaction. Measurable predictions are robust against 
shifting the classical--quantum boundary (von Neumann--cut).}

\section{Introduction}
The standard quantum theory, as it is described e.g. in Ref.\cite{Per.Boo},
can be analyzed from naive dynamic aspects. It turns out of course that the
interaction of classical and quantum degrees of freedom are treated 
asymmetrically and also non-dynamically. This issue is exposed in the 
Sect.~2. While Copenhagen quantum mechanics is a correct theory as it stands,
I propose a dynamical formulation which is thought to be equivalent to it. 
In Sec.~3, I present certain 'hybrid' differential equations of motion, 
developed in several early\cite{Dio.Two,Dio.True,Dio.Opt,Dio.Ovi} and 
recent\cite{Roy,Dio.Bie} works, profiting from related
ideas\cite{SheSud,Ale,BouTra,Sal} as well. The Sec.~4 is devoted to a proof
that hybrid dynamics in proper limit reproduces the system--detector 
interaction exactly in the form as it is postulated within the collapse 
theory\cite{Neu} of standard quantum mechanics. In Sec.~5, I verify that, 
similarly to the standard theory, the hybrid dynamics experimental predictions 
are robust against the shift of the boundary (von Neumann cut) between 
classical and quantized systems. Finally, an outlook is given in Sec.~6.

\section{Copenhagen Universe: $C\times Q$}
The physical world of standard (Copenhagen) quantum mechanics is dichotomic.
The Universe $U$ consists of a classical and a quantized part: 
\be
U=C\times Q.
\ee
Macroscopic degrees of freedom, building up $C$, are invariably described by 
classical {\it canonical\/} equations:
\be
\frac{d x}{dt}=\p_p H_C(x,p),~~~\frac{d p}{dt}=-\p_x H_C(x,p),
\ee
while microscopic degrees of freedom $(Q)$ are quantized, their quantum state 
$\hat\r_Q$ obeys quantum dynamics 
\be
\frac{d\hat\rho_Q}{dt}=-i[\hat H_Q(x,p),\hat\r_Q]~,
\ee
where the Hamiltonian operator depends on the classical dynamic variables of 
$C$ as external parameters. The macroscopic variables $x,p$ are usually 
{\it robust\/} variables as compared to the quantized microscopic variables. 
Therefore, the {\it back-reaction} of the quantized system upon the classical
one can and used to be ignored. We just keep going on with Eqs.~(2).

There is a paradigmatic exception, however. There are macroscopic systems which
are not robust against back-reactions from certain microsystems. Instead, these
macrosystems are definitely sensitive and will magnify the tiniest 
back-reaction. As a rule, we use such macrosystems as {\it detectors\/} 
detecting the given quantized microsystem. The process of back-reaction is 
called the quantum measurement process. Standard Copenhagen theory does not 
intend to describe the details of such measurement process. It considers 
idealized cases of measurement where, instead of solving dynamic equations, 
only the random final states $\hat\r^f$ after the back-reaction (measurement) 
are specified in terms of the initial state $\hat\r^i$ (before measurement):
\be
\hat\r^f=\cases{
\frac{1}{p_1}\hat P_1\hat\r^i\hat P_1&
~~~with~probability~$p_1=tr[\hat P_1\hat\r^i]$\cr
\frac{1}{p_2}\hat P_2\hat\r^i\hat P_2&
~~~with~probability~$p_2=tr[\hat P_2\hat\r^i]$\cr
~~~\dots\cr
\frac{1}{p_n}\hat P_n\hat\r^i\hat P_n&
~~~with~probability~$p_n=tr[\hat P_n\hat\r^i]$\cr
~~~\dots}
\ee
where the $\hat P_n$'s form a complete orthogonal set of Hermitian projectors.

As for the back-reaction of the quantized system on the detector's classical
variables, it is tacitly understood that one of them, ideally the position
(say $x$) of the detector's pointer, taking a neutral value $x^i$ before, will
become completely correlated with the random set (4) of final quantum states 
$\hat\r^f$: 
\be
x^f\approx gn,~~~~\hbox{with probability}~p_n,~~~n=1,2,\dots
\ee
The detection process (4,5) is usually considered as the measurement of the
Hermitian observable $g\hat A\equiv g\sum_n n\hat P_n$, the random outcomes (5) 
are the eigenvalues of $g\hat A$.

The interaction between 'detectors' and quantum systems, as specified by the 
standard Eqs.~(4,5), differs from the Eq.~(3) which is valid otherwise. But this
is just another appearance of the basic dichotomy, without endangering logical
consistency of the standard theory. Nonetheless, we present a certain 'hybrid' 
dynamics which treats all interactions by the same differential equations.

The consistency of the standard theory relies heavily upon the robustness of its 
measurable predictions against shifting the boundary between the classical and 
the quantized parts of the Universe:
\be
C\times Q \rightarrow C'\times Q'~.
\ee
I mention, however, the notorious extreme case\cite{Eve} $U=Q$ where the 
standard quantum theory becomes paralyzed by the impossibility of detection due 
to the complete lack of classical systems\cite{Dio.Kop}. Still, pushing the 
boundary toward more and more quantized degrees of freedom is necessary since we 
have no evidence of an upper limit for the size of the quantized system $Q$. The
back-reaction of the enlarged quantum system may grow significant. This would 
also motivate the next Section.

\section{Hybrid dynamics for $C\times Q$}
Constructing unified dynamics of hybrid systems requires compromises, c.f.
Salcedo's no-go proofs\cite{Sal}. Sharp values of the classical variables $x,p$, 
as well as sharply given phase space trajectories must be given up if the 
classical system $C$ is brought into interaction with the quantized system $Q$. 
Classical states will be described by phase space distributions $\r_C(x,p)$
which should never be narrower than a single Planck cell neither should it bear 
particular structures within single Planck cells.

If the classical and quantized systems are uncorrelated then it is 
straightforward to construct the hybrid state
\be
\hat\r(x,p)=\hat\r_Q\r_C(x,p)
\ee
for the composite system $C\times Q$. In general, we represent the state of the 
hybrid system by a hybrid 'density' $\hat\r(x,p)$ which is a phase space 
dependent non--negative operator. Its trace is the phase space distribution 
$\r_C(x,p)$ of $C$ while its phase space integral yields the density operator 
$\hat\r_Q$ of $Q$. When $\hat\r(x,p)$ is not factorable the unconditional 
quantum state $\hat\r_Q$ must be distinguished from the conditional quantum 
states:
\be
\hat\r_{xp}=\frac{\hat\r(x,p)}{\r_C(x,p)}
\ee
depending on the classical coordinates $x,p$ as conditions. The generic hybrid 
observables are Hermitian operators $\hat F(x,p)$ depending on the classical 
variables as well. Their expectation values are defined as follows:
\be
\left\langle \hat F(x,p) \right\rangle = tr\int\hat F(x,p)\hat\r(x,p)dxdp.
\ee
The Hamiltonian of the hybrid system takes this form:
\be
\hat H(x,p)=H_C(x,p) +\hat H_Q(x,p)~.
\ee
One can construct the following canonical hybrid equation of motion for the 
hybrid state $\hat\r(x,p)$:
\be
\frac{\hat\r(x,p)}{dt}= 
-i:\!\!\hat H\left(x+{\p_x+i\p_p\over2},p+{\p_p-i\p_x\over2}\right)\!\!:
\hat\r(x,p) + H.C. 
\ee
where $:\dots:$ mean that all differentiations must be done first. There is an 
additional analytic condition for $\hat\r(x,p)$ which assures the mathematical 
consistency of the hybrid equation\cite{Roy}. 

If all classical variables were robust then Eq.~(11) would reduce asymptotically 
to the standard Eqs.~(2,3). In the next Section, however, we consider the 
paradigmatic opposite case: system--detector interaction.

\section{Measurement interaction with hybrid dynamics}
We can describe the schematic interaction (4,5) between the classical detector 
and the quantized system as a hybrid dynamic process. Let the detector's pointer 
be a harmonic oscillator with Hamiltonian $\half(x^2+p^2)$, shortly but strongly 
coupled to the observed quantum variable $g\hat A=g\sum_n n\hat P_n$ by the 
following interaction Hamiltonian: 
\be
\hat H_Q(x,p)=\delta(t)pg\hat A~.
\ee
Since $1/g$ will be the relative precision of the measurement we assume $g\gg1$. 
We are interested in the states just before and, respectively, after the 
measurement hence the interaction Hamiltonian (12) dominates and the hybrid 
equation (11) will take this form:
\be 
\frac{d\hat\r}{dt}=
-ig\delta(t)p[\hat A,\hat\r]-\frac{1}{2}g\delta(t)[\hat A,\p_x\hat\r]_{+}
-\frac{i}{2}g\delta(t)[\hat A,\p_p\hat\r]~.
\ee
As it follows from this dynamics, $x$ will play the role of the pointer variable 
to indicate the value $g,2g,\dots$ of the operator $g\hat A$ after the 
measurement. We assume the following factorized initial state for the hybrid 
system:
\be
\hat\r^i\frac{\exp[-\half(x^2+p^2)]}{2\pi}\equiv\hat\r^i\r_C^i(x,p)~,
\ee
where $\r_C^i(x,p)$ corresponds to the pointer's initial position $x^i=0\pm1$.
The evolution (13) acts on block-matrix elements of the initial state (14) as 
follows:
\begin{eqnarray}
&&\exp\Bigl(-igp[\hat A~,~.]-\frac{1}{2}g[\hat A\p_x~,~.]_{+}
           -\frac{i}{2}g[\hat A\p_p~,~.]\Bigr)
\hat P_n\hat\r^i\hat P_m \r_C^i(x,p)=\nonumber\\
&=&\exp\left(-\frac{(n-m)^2}{8}g^2+\frac{n-m}{2i}gp\right)
\hat P_n\hat\r^i\hat P_m\r_C^i\left(x-\frac{n+m}{2}g,p\right)~. 
\end{eqnarray}
Since $g\gg1$ the off--diagonal terms are heavily damped, so the initial hybrid
state (14) tends into the diagonal final state:
\be
\sum_n \hat P_n\hat\r^i\hat P_n \r_C^i(x-n g,p)~.
\ee
This result clearly shows that the pointer's coordinate shifts to the 
statistically well--separated positions $x^f=gn\pm1$, $(n=1,2,\dots)$ with 
probability $p_n=tr[\hat P_n\hat\r^i]$. The quantized system's conditional 
state (8) is $\hat P_n\hat\r^i\hat P_n/p_n$, respectively. This scheme of the 
final quantum state $\hat\r^f$ {\it and} classical pointer position $x^f$ is 
identical to the result (4,5) of the standard theory.
 
\section{Shifting the quantum-classical boundary}
I am going to prove the consistency of hybrid dynamics against the shift (6) of 
quantum {\it vs.\/} classical boundary. The new hybrid states, Hamiltonian, and 
observables will be denoted by $\hat\r'(x',p'), \hat H'(x',p')$ and 
$\hat F'(x',p')$, respectively. First, we have to establish their correspondence
with the old ones. Without restricting generality, one can consider a simplest
shift in favor of $Q$ on $C$'s account. We quantize $x_1,p_1$, the other ones 
$(x_2,p_2,x_3,p_3,\dots)\equiv(x',p')$ will remain classical. Then the new 
hybrid Hamiltonian should be chosen as
\be
\hat H'(x',p')=:\!\!\hat H(\hat x_1,\hat p_1,x',p')\!\!:
\ee
where $:\dots:$ means normal ordering for $\hat x_1,\hat p_1$. The new hybrid 
state must be defined by the implicit relation
\be
\langle x_1,p_1\vert\hat\r'(x',p')\vert x_1,p_1\rangle=\hat\r(x,p)
\ee
where $\vert x_1,p_1\rangle$ are normalized coherent states, i.e. eigenstates 
of $\hat x_1+i\hat p_1$. This relation defines $\hat\r'(x',p')$ uniquely since, 
as we mentioned in Sec.~3, a certain analyticity condition has been imposed on 
$\hat\r(x,p)$. Now assume that $\hat\r(x,p)$ satisfies the hybrid Eq.~(11). Then 
it follows from the construction\cite{Dio.Ovi,Roy} of the hybrid dynamics that 
the shifted hybrid state (18) will satisfy the hybrid equation (11) with the 
shifted Hamiltonian (17). 

We have to prove that the physical predictions are robust against the shift. Let 
us construct the shifted observable $\hat F'(x',p')$ from the original one in 
the following way:
\be
\hat F'(x',p')=\{\hat F(\hat x_1,\hat p_1,x',p')\}_{sym}
\ee
where a Wigner-Weyl symmetrization for $\hat x_1,\hat p_1$ is understood on the 
r.h.s. If $\hat F(x,p)$ is a smooth enough function of $x_1,p_1$ on the scale of 
Planck cell then the expectation value of $\hat F'(x',p')$ in state 
$\hat\r'(x',p')$ tends to be identical to the expectation value of $\hat F(x,p)$ 
in $\hat\r(x,p)$. (The exact relation between the two expectation values is 
given in Refs.\cite{Dio.Opt,Dio.Ovi}.) This assures the robustness of physical 
predictions provided the observables are smooth functions {\it vs.\/} the Planck 
cell.

We see that the shift of von Neumann's cut in one direction means quantization 
of a classical canonical system. In the opposite direction it means 
de-quantization of a quantized system. These two procedures are exact inverses 
of each other. As it is well--known in standard quantum theory, canonical 
quantization (hence de-quantization, too) can be defined in many ways, depending 
e.g. on the operator ordering of the Hamiltonian. The concrete meaning of the 
shift of von Neumann's cut depends on conventions of 
quantization/de-quantization. Hay and Peres\cite{HayPer} have recently applied 
Wigner--functions as the phase--space distribution of de-quantized systems. In 
our hybrid dynamics the Husimi--function (18) plays the same role. Hay and 
Peres suggest the invariance of cascaded standard measurements against the shift 
of von Neumann's cut\footnote{
Such invariance has been pointed out within a phenomenological theory of 
open systems\cite{SDG}, intrinsically related to hybrid dynamic theory.}
provided the quantized pointer variables are Wigner--Weyl--ordered Hermitian 
operators. In hybrid dynamics, however, these pointer variables should be the 
normal--ordered ones (17). The discrepancy follows from the different 
conventions of quantization.

In fact, in hybrid dynamics we departed from the more popular convention of
classical--quantum correspondence and we have replaced Wigner--functions by
Husimi--functions. This allows de-quantizing any quantum states including
those with non-positive Wigner--functions. The price we pay is the loss of
reversibility of hybrid dynamics. On the other hand, it can thus describe 
the standard Copenhagen system--detector interaction which {\it is\/} 
irreversible.

\section{Outlook}
The introduction of hybrid dynamics is not a necessity, it is a possibility. It 
does not alter the physical predictions of standard quantum theory. It is 
thought to reproduce them. Hybrid dynamics turns out to be surprisingly
effective in cases of 'continuous' observation' or massive continuous
'back-reaction' of quantized systems where standard detector concept requires 
cumbersome considerations (though they work, finally) or gross simplifications 
(which are always restrictive). These positive features of hybrid dynamic 
approach have recently been emphasized\cite{Roy} by the present author, Gisin, 
and Strunz. In particular, I do expect relevant applications of hybrid dynamics 
in quantum cosmology where massive 'back-reaction' is being currently treated by 
a number of semi-phenomenological models\cite{HuMat,Hal}.

\section*{Acknowledgments}
I am indebted to the conference organizers, especially to H.-D. Doebner, for the
invitation and for the generous support. My research was supported in part by 
the Hungarian Science Research Fund under contract T 016047.

\end{document}